\documentclass[ %
 reprint,
 amsmath,amssymb,
 aps,
]{revtex4-2}
\usepackage{xcolor}
\usepackage{graphicx}
\usepackage{dcolumn}
\usepackage{bm}

\begin{document}
\preprint{RBI-ThPhys-2025-06}

\title{\mbox{Corrections to Kerr-Newman black hole from noncommutative
Einstein-Maxwell equation}}
\author{Filip Požar}
\email{filip.pozar@irb.hr}
\affiliation{Rudjer Boskovic Institute, Bijenička c.54, 10002-Zagreb, Croatia
}%

\date{\today}

\begin{abstract}
In this letter we introduce the noncommutative geometry into the standard Einstein-Hilbert-Maxwell action via the $\partial_t\wedge\partial_\varphi$ Drinfeld twist and solve the equation of motion pertubatively in the expansion of the noncommutative parameter $a$. The equation of motion, the NC Einstein-Maxwell equation, turns out to be effectively a problem in nonlinear electrodynamics where the energy-momentum tensor $T_{\mu\nu}$ obtains correction terms with three Faraday tensors $F_{\mu\nu}$. A solution with nonzero $a^1$ terms turns out to be the Kerr-Newman black hole modified with nonzero $g_{t\theta}, g_{r\varphi}, g_{tr}$ and $g_{\varphi\theta}$ components proportional to $a$, while the electromagnetic potential is the Seiberg-Witten expanded Kerr-Newman potential which introduces a nonzero $A_{\theta}$ term proportional to $a$.
\end{abstract}

\maketitle


\section{\label{sec:level1}Introduction}
Noncommutative (NC) geometry is a candidate framework for probing the effects of quantum spacetime in physics. It has been studied in many papers, concerning various topics like black hole entropy \cite{Gupta:2022oel, Juric:2022bnm, Hrelja:2024tgj}, gravitational waves \cite{Herceg:2023pmc, Zhao:2023uam}, gauge theory \cite{Ciric:2017rnf,DimitrijevicCiric:2024ibc, Maris:2024fvp}, vacuum energy \cite{Mignemi:2017yhd, Juric:2018qdi},  etc.. In this letter, we will use the Hopf algebra formalism, as outlined in \cite{Aschieri:2009qh,Juric:2022bnm, Herceg:2023pmc} and many other references, to deform the Einstein-Hilbert-Maxwell (EHM) action and solve its equations of motion under the assumption that the solution fields have two Killing vectors, $\partial_t$ and $\partial_\varphi$. We conclude that, in the $\partial_t\wedge \partial_\varphi$ twist, there exist infinitely many deformations of the EHM action, but all of them share a static solution which we found.\\

This investigation is related to the current topics in noncommutative geometry in the sense that it uses the same Drinfeld twist approach to noncommutative geometry as the recent papers \cite{Gupta:2022oel, Juric:2022bnm, Hrelja:2024tgj, Herceg:2023pmc, Zhao:2023uam, Ciric:2017rnf,DimitrijevicCiric:2024ibc, Aschieri:2009qh} (and many more) as well as the well established Seiberg-Witten map approach to noncommutative gauge theory \cite{Seiberg:1999vs}, but offering an inspection of new, previously unconsidered, phenomenological qualities. Namely, this is the first theoretical observation of a corrected black hole metric arising from the effects of noncommutative geometry as the solution of the equation of motion for some NC action written in NC geometry. Additionally, this is the first investigation of the Kerr-Newman black hole in this framework. The corrections obtained are significant both as a proof of concept, demonstrating that physics in noncommutative geometry can ambitiously pursue nontrivial solutions to the noncommutative Einstein equation, and from a phenomenological perspective, since the corrected metric governs all gravitational properties within the solved system.\\

In short, the (commutative) geometry of a (commutative) manifold $\mathcal{M}$ is encoded in its Hopf Algebra of vector fields $(\mathcal{U}(\Xi),m, \Delta, \epsilon, S)$, where
\begin{itemize}
    \item $\mathcal{U}(\Xi)$ is the universal enveloping algebra of the vector field Lie algebra $\Xi$. It is a unital associative algebra.
    \item $m:\mathcal{U}(\Xi)\otimes \mathcal{U}(\Xi) \rightarrow \mathcal{U}(\Xi)$ is the canonical product of the $\mathcal{U}(\Xi)$ Hopf algebra, given as the composition of vector fields.
    \item $\Delta:\mathcal{U}(\Xi)\rightarrow \mathcal{U}(\Xi)\otimes \mathcal{U}(\Xi)$ is the canonical coproduct defined as $\Delta(v) = v\otimes 1 + 1\otimes v$, for all $v\in \mathcal{U}(\Xi)$.
    \item $\epsilon : \mathcal{U}(\Xi)\rightarrow \mathbb{C}$ is the canonical counit defined as $\epsilon(v) = 0$, for all $v\in \mathcal{U}(\Xi)$.
    \item $S:\mathcal{U}(\Xi) \rightarrow \mathcal{U}(\Xi)$ is the canonical $\mathcal{U}(\Xi)$ anti-automorphism defined as $S(v) = -v$, for all $v\in \mathcal{U}(\Xi)$.
\end{itemize}
Together those 5 objects satisfy various structural identities which define an abstract Hopf algebra. It is known that using a Drinfeld element \footnote{which satisfies the technical cocycle and normalization conditions} $\mathcal{F}\in \mathcal{U}(\Xi)\otimes \mathcal{U}(\Xi)$ it is possible to systematically deform the starting Hopf algebra into a new, deformed, one $(\mathcal{U}(\Xi), m, \Delta_\mathcal{F}, \epsilon, S_\mathcal{F})$ with
\begin{equation}
    \begin{split}
        \Delta_\mathcal{F} &= \mathcal{F}\Delta \mathcal{F}^{-1}\;,\\
        S_\mathcal{F} &= \alpha S \alpha^{-1},\quad \alpha = m\left[ \mathcal{F}\;1\otimes S\right]\;,
    \end{split}
\end{equation}
and this deformed Hopf algebra is said to encode the noncommutative geometry of some noncommutative manifold $\mathcal{M}_\star$. Furthermore, the algebra modules, i.e., the algebras on which the Hopf algebra acts \footnote{All of the relevant geometric modules on which the vector field Hopf algebra naturally acts can be seen as algebras - e.g., the pointwise algebra of functions on the manifold $C^\infty(\mathcal{M},\mathbb{C})$, the tensor algebra $\mathcal{T}$, etc. are all algebras.}, of the twisted Hopf algebra are well known. Any algebra $(A,\star)$ module of the deformed Hopf algebra can be obtained from the underlying undeformed Hopf algebra's module $(A,\cdot)$ simply by setting \footnote{Where $a_1 \cdot a_2$ can be understood as a tensor product map $\cdot : A\otimes A \rightarrow A$ with $\cdot(a_1\otimes a_2) = a_1\cdot a_2$.}
\begin{equation}
    a_1\star a_2 = \cdot\circ \mathcal{F}^{-1}\left(a_1\otimes a_2\right)\;,\quad\forall a_1,a_2\in A\;.
\label{star product}
\end{equation}
With the definition \eqref{star product} it is easy to see that $(A,\star)$ is an algebra module of the deformed Hopf algebra because
\begin{equation}
    v\vartriangleright (a_1\star a_2) = \star \left[\Delta_\mathcal{F}(v)\vartriangleright \left(a_1\otimes a_2\right)\right]
\label{deformed action}
\end{equation}
holds for all $v\in \mathcal{U}(\Xi)$ and all $a_1,a_2 \in A$. We will use the construction \eqref{star product} and its consequence \eqref{deformed action} to construct an action functional that is covariant under the deformed Hopf algebra obtained by deforming the $\mathcal{U}(\Xi)$ Hopf algebra with the $\partial_t\wedge\partial_\varphi$ Drinfeld twist
\begin{equation}
    \mathcal{F} = e^{-i\frac{a}{2}\left(\partial_t \otimes \partial_\varphi - \partial_\varphi \otimes \partial_t \right)}=e^{-i\frac{a}{2} \partial_t \wedge \partial_\varphi}\;.
\label{drinfeld twist}
\end{equation}
The Drinfeld twist \eqref{drinfeld twist} is of the Moyal type (and is also Abelian) which has the general form
\begin{equation}
    \mathcal{F} = e^{-\frac{i}{2}\theta^{\mu\nu}\partial_\mu \otimes \partial_\nu}\;,
\label{moyal}
\end{equation}
with the tensor $\theta$ being constant and antisymmetric.

\section{Noncommutative Einstein-Hilbert-Maxwell action}
In this section we will obtain the noncommutative Einstein-Hilbert-Maxwell action. Let us consider the deformed Hopf algebra $(\mathcal{U}(\Xi), m, \Delta_\mathcal{F}, \epsilon, S_\mathcal{F})$ of vector fields over spacetime $\mathcal{M}$ and its deformed algebra module of spacetime functions $C^\infty(\mathcal{M},\mathbb{C})$, defined using the Drinfeld twist \eqref{drinfeld twist} as above. An example action in this framework is, e.g., the NC charged scalar field and electromagnetism action on the fixed Reissner–Nordström (RN) background defined in \cite{Ciric:2017rnf}:
\begin{equation}
\begin{split}
S = \int d^4x \sqrt{-g}&\left[g^{\mu\nu}\cdot \hat{D}_\mu \hat{\phi}^\dagger \star \hat{D}_\nu \hat{\phi}  - \mu^2 \hat{\phi}^\dagger \star \hat{\phi}\right. \\
&\left. -\frac{1}{4}g^{\mu\nu}\cdot g^{\rho\sigma}\cdot \hat{F}_{\mu\rho}\star \hat{F}_{\nu\sigma} \right]\;,
\label{Andelo action}
\end{split}
\end{equation}
with the deformed gauge covariant derivative defined as
\begin{equation}
    \hat{D}_\mu \hat{\phi} = \partial_\mu \hat{\phi} - iq\hat{A}_\mu \star \hat{\phi}\;,
\end{equation}
and with the star product
\begin{equation}
\begin{split}
\left(f\star g\right)(x) =& \cdot\left[\mathcal{F}^{-1} \left(f\otimes g\right)\right] \\
=&\cdot\left[e^{i\frac{a}{2}\partial_t\wedge \partial_\varphi}\left(f\otimes g\right)\right]\\
=& f(x)\cdot g(x) + i\frac{a}{2}\left[\partial_tf(x)\partial_\varphi g(x) \right. \\
&\left. - \partial_\varphi f(x)\partial_t g(x)\right] + \mathcal{O}(a^2)\;,
\end{split}
\label{star product tphi}
\end{equation}
which is a direct application of \eqref{star product} to the Drinfeld twist \eqref{drinfeld twist}.
It is important to comment, because it will be key to including the Einstein-Hilbert Lagrangian into the action \eqref{Andelo action}, that in \eqref{Andelo action} the $\star$ product between metric tensors was replaced with a standard commutative product. That is because $\partial_t$ and $\partial_\varphi$ are Killing vectors of the RN background and as such any function of the metric obeys
\begin{equation}
    F(g_{\mu\nu})\star G(x) = F(g_{\mu\nu})\cdot G(x)
\label{FG property}
\end{equation}
due to \eqref{star product tphi}. Additionally, since the fields in \eqref{Andelo action} are not elements of the algebra $(C^\infty(\mathcal{M},\mathbb{C}),\cdot)$, but of the deformed algebra $(C^\infty(\mathcal{M},\mathbb{C}),\star)$, the usual commutative $U(1)$ gauge symmetry is not a symmetry of \eqref{Andelo action}. The only type of gauge transformation that the action \eqref{Andelo action} can be invariant to is the noncommutative $U(1)$ gauge transformation, denoted as $U(1)_\star$, whose infinitesimal expansion is given as
\begin{equation}
\begin{split}
\delta_{\hat{\lambda}}\hat{\phi} &= \hat{\lambda}\star \hat{\phi}\;, \\
\delta_{\hat{\lambda}}\hat{A}_\mu &= \partial_\mu \hat{\lambda} + i\left(\hat{\lambda}\star \hat{A}_\mu - \hat{A}_\mu \star \hat{\lambda} \right)\;,
\end{split}
\end{equation}
with $\hat{\lambda}(x)$ being the NC gauged parameter. To distinguish the fields obeying NC gauge transformations with the fields obeying commutative gauge transformations, it is customary to denote NC gauge theory fields with a hat, e.g., $\hat{A}_\mu$, and the fields obeying usual $U(1)$ transfromation laws without the hat, e.g., $A_\mu$. In fact, the action \eqref{Andelo action} is symmetric to $U(1)_\star$ gauge transformations.\\

One can see that in the commutative limit $a\rightarrow 0$ the theory given by the action \eqref{Andelo action} becomes the usual electromagnetism + charged scalar field theory in the RN background and the $U(1)_\star$ gauge symmetry of the action reverts back to $U(1)$ gauge symmetry. A useful concept in the field of noncommutative gauge theory is the Seiberg-Witten (SW) map \cite{Seiberg:1999vs} which maps a commutative gauge transformation $(\lambda, A_\mu, \phi)$ into a NC (with respect to the Moyal twist \eqref{moyal}) one $(\hat{\lambda}, \hat{A}_\mu, \hat{\phi})$ using the identification
\begin{equation}
\begin{split}
\hat{\lambda} &= \lambda + \frac{q}{2}\theta^{\mu\nu}\partial_\mu \lambda \cdot A_\nu + \mathcal{O}(a^2)\;,\\
\hat{A}_\mu &= A_\mu -\frac{q}{2} \theta^{\rho\sigma}A_\rho\left(\partial_\sigma A_\mu + F_{\sigma\mu}\right) + \mathcal{O}(a^2)\;,\\
\hat{\phi} &= \phi - \frac{q}{4}\theta^{\rho\sigma}A_\rho\left(\partial_\sigma \phi + D_\sigma \phi\right) + \mathcal{O}(a^2)\;,
\end{split}
\label{SW map}
\end{equation}
where $q$ is the electric charge of the scalar field $\phi$.
By using the Seiberg-Witten map \eqref{SW map} and expanding the star product \eqref{star product tphi} up to first order, the authors of \cite{Ciric:2017rnf} were able to express the action \eqref{Andelo action} completely in terms of commutative fields and commutative products. The purely electromagnetic sector of the expanded action equals
\begin{equation}
\begin{split}
S \supset \int d^4x\sqrt{-g}&\Biggl[-\frac{1}{4}F_{\mu\nu}F^{\mu\nu} +\frac{q}{8}\theta^{\alpha\beta}\Bigl( F_{\alpha\beta}F_{\mu\nu}F^{\mu\nu}\\
& - 4F_{\mu\alpha}F_{\nu\beta}F^{\mu\nu}\Bigr)\Biggr]+ \mathcal{O}(a^2)\;.
\end{split}
\label{Andelo lagrangian}
\end{equation}
We will obtain a similar expression for the Einstein-Hilbert-Maxwell Lagrangian.
\\

If noncommutative geometry is a valid description of gravity coupled to classical electromagnetism in the regimes where gravity is not just a background, then the action in this case needs to have the form
\begin{equation}
    S[g,\hat{A}] = \int d^4 x \left[\mathcal{L}^\star_G(g) + \mathcal{L}^\star_{EM}(g,\hat{A}) \right]\;,
\label{NC action}
\end{equation}
where $\mathcal{L}^\star_G$ and $\mathcal{L}^\star_{EM}$ are the NC Lagrangian densities for the gravitational and electromagnetic sector and they need to reproduce the usual Einstein-Hilbert and Maxwell Lagrangian densities in the commutative limit.\\

In this paper, we consider only the minimally NC deformed Lagrangians which only replace pointwise commutative $\cdot$ products with $\star$ products and gauge fields with Seiberg-Witten expanded NC gauge fields in the Einstein-Hilbert-Maxwell Lagrangian density. The problem is that there are infinitely many such densities. For example, $\mathcal{L}^{\star (1)}_{EM}$, $\mathcal{L}^{\star (2)}_{EM}$ and $\mathcal{L}^{\star (3)}_{EM}$
\begin{equation}
\begin{split}
\mathcal{L}^{\star (1)}_{EM} &= \sqrt{-g}\star g^{\mu\nu}\star g^{\rho\sigma}\star \hat{F}_{\mu\rho}\star \hat{F}_{\nu\sigma}\; \\
\mathcal{L}^{\star (2)}_{EM} &= \sqrt{-g}\star g^{\mu\nu}\star \hat{F}_{\mu\rho}\star g^{\rho\sigma}\star \hat{F}_{\nu\sigma}\;\\
\mathcal{L}^{\star (3)}_{EM} &= -2\mathcal{L}^{\star (1)}_{EM} + 3\mathcal{L}^{\star (2)}_{EM}
\end{split}
\label{deformations}
\end{equation}
all reproduce the Maxwell Lagrangian in the commutative limit, but they all give rise to mutually inequivalent action functionals. When talking about NC deformations of a theory, the author means choosing one such NC Einstein-Hilbert-Maxwell Lagrangian arising from the twist \eqref{drinfeld twist} and the NC geometry and gauge theory frameworks outlined in Sections I and II \footnote{One can imagine also adding some fundamentally new terms to the Einstein-Hilbert-Maxwell action proportional to the NC parameter $a$. For example, $a R^2$ or almost anything else. Such actions would also revert back to the standard action in the commutative limit. But, in this paper we only consider the deformations which arise from submerging the Einstein-Hilbert-Maxwell action into the NC geometry and NC gauge theory framework without adding any new terms.}. From now on, we will suppose that \eqref{NC action} are minimal NC deformations in the sense that was just described. So far, the author is not aware of any criterion to prefer any one NC Lagrangian density over all other candidates. A similar problem is present in the NC gauge theory of gravity \cite{Juric:2025kjl} where infinitely many noncommutative generalizations of geometric quantities, e.g, the NC metric, curvature scalars etc., have the correct commutative limits, but it is impossible to prefer any one choice over the others. But, it is still possible to find a solution to the equations of motion of the action \eqref{NC action}. Namely, since all of the NC corrections to the action \eqref{NC action} come from the star product \eqref{star product tphi} and from the SW map \eqref{SW map}, a solution $(g, A)$ of the expanded action whose fields have Killing vectors $\partial_t$ and $\partial_\varphi$, has to be the same as the solution of the action
\begin{gather}
S^{\text{NC}}_{\text{EHM}}[g,A] = \int d^4x \sqrt{-g}\left[ R -\frac{1}{4} g^{\mu\nu}g^{\rho\sigma}\hat{F}_{\mu\rho}\cdot\hat{F}_{\nu\sigma} \right]\label{NC EHM}\\
=\int d^4x\sqrt{-g}\biggl[R -\frac{1}{4}F_{\mu\nu}F^{\mu\nu} -\frac{q}{2}\theta^{\alpha\beta} F_{\mu\alpha}F_{\nu\beta}F^{\mu\nu}\biggr] +\mathcal{O}(a^2)\notag
\end{gather}
with the same kind of symmetry in $g$ and $A$. This is due to the property \eqref{FG property} applied to $A$ and $g$. The action \eqref{NC EHM}, which we refer to as the NC Einstein-Hilbert-Maxwell (NC EHM) action, can be thought of as the restriction of the action functional \eqref{NC action} to the part of the classical configuration space where the fields have Killing vectors $\partial_t$ and $\partial_\varphi$. Additionally, the electromagnetic part of the action \eqref{NC EHM} expands to \eqref{Andelo lagrangian} (with $F_{\alpha\beta}$ terms vanishing due to the time and axial symmetry), so \eqref{NC EHM} is a nonlinear electrodynamics theory whose nonlinearity is induced by spacetime noncommutativity/quantumness. In conclusion, in a special symmetrical regime it is possible to actually determine the form of the NC action \eqref{NC action} without specifying the ordering of $\star$ products. Now it is just a question whether a solution with $\partial_t$ and $\partial_\varphi$ Killing vectors exists for NC EHM action \eqref{NC EHM}. The answer is that it does and we will find it in the following sections.\\

\section{The NC Einstein-Maxwell equation}
Having obtained the noncommutative Einstein-Hilbert-Maxwell action, we can vary it and obtain the equations of motion for $g_{\mu\nu}$ and $A_\mu$, keeping in mind that any solution that we find has to obey the condition that $\partial_t$ and $\partial_\varphi$ are Killing vectors for the metric and for the potential. The equation of motion for the metric, up to $a^1$ order, is the NC Einstein equation
\begin{equation}
    G_{\mu\nu} = 2T_{\mu\nu}\;,
\label{EFE}
\end{equation}
with
\begin{equation}
\begin{split}
    &T_{\mu\nu} = \frac{1}{4} g_{\mu\nu} F_{\mu\nu}F^{\mu\nu} + F_{\mu}\,^\rho F_{\nu\rho} -\\
    &\frac{q}{2}\theta^{\alpha\beta}\left(F_{\mu\alpha}F_{\sigma\beta}F_{\nu}\,^\sigma + F_{\nu\alpha}F_{\sigma\beta}F_{\mu}\,^\sigma\right)+ \mathcal{O}(a^2)\;.
\end{split}
\label{Tmunu}
\end{equation}
On the other hand, the NC Maxwell equation is given in \cite{Ciric:2017rnf}, but in our case the metric is dynamical instead of the Reissner Nordstrom background:
\begin{equation}
\begin{split}
&\partial_\mu F^{\mu\lambda} + \Gamma^{\rho}_{\mu\rho} F^{\mu\lambda} + q\theta^{\alpha\beta}\biggl(-\frac{1}{2}\partial_\mu\left(F_{\alpha\beta} F^{\mu\lambda}\right)\\
&-\frac{1}{2} \Gamma^{\rho}_{\mu\rho}F_{\alpha\beta}F^{\mu\lambda} + \partial_\mu \left(F_\alpha\,^\mu F_\beta\;^\lambda \right) -\\
&\partial_\alpha\left(F_{\beta_\mu}F^{\mu\lambda}\right)\biggr) + q\theta^{\alpha \lambda}\biggl(\frac{1}{2}\partial_\mu\left(F_{\alpha\nu}F^{\mu\nu}\right)+\\
&\frac{1}{2}\Gamma^{\rho}_{\mu\rho}F_{\alpha\nu}F^{\mu\nu} - \frac{1}{4}\partial_\alpha\left(F_{\mu\nu}F^{\mu\nu}\right)\biggr) + \mathcal{O}(a^2) = 0^\lambda\;.
\end{split}
\label{Maxwell}
\end{equation}
We will approach solving \eqref{EFE} and \eqref{Maxwell} perturbatively in an expansion in $a$, disregarding $a^2$ effects. From the form of the energy-momentum tensor \eqref{Tmunu} and of the Maxwell equation \eqref{Maxwell}, we can see that the commutative solutions to the Einstein-Maxwell problem do not solve the NC problem \footnote{Unless some accidental cancellation of seemingly unrelated terms in \eqref{EFE} and \eqref{Maxwell} happens which would keep the equations of motion robust to noncommutative corrections. That is not the case for Kerr-Newmann perturbations and there remain nonzero $a^1$ terms which are not solved with the purely commutative solution.} in case the electromagnetic potential $A_\mu$ has $t$ and $\varphi$ components, i.e., the electrovacua which describe spaces with both nonzero magnetic and electric fields are described with modified equations of motion. Having said that, we will search for a solution $(g,A)$ which is a perturbation of the Kerr-Newmann (KN) black hole (without assuming slow rotation)
\begin{equation}
\begin{split}
g_{\mu\nu} &= g_{\mu\nu}^{\text{KN}} + ah_{\mu\nu}\\
A_{\mu} &= A_\mu^{\text{KN}} + a B_\mu\;,
\label{ansatz}
\end{split}
\end{equation}
since the Kerr-Newmann commutative solution is such that the Einstein and Maxwell equations will have nonzero $a^1$ terms - which we can try to find a nontrivial solution for in terms of $h_{\mu\nu}$ and $B_\mu$. Plugging in the Kerr-Newmann solution
\widetext
\begin{equation}
\begin{gathered}
g^{\text{KN}}_{\mu\nu} = \left[\begin{matrix}\frac{\frac{J^{2} \sin^{2}{\left(\theta \right)}}{M^{2}} - \Delta{\left(r \right)}}{\Sigma{\left(r,\theta \right)}} & 0 & 0 & - \frac{J \left(\frac{J^{2}}{M^{2}} + r^{2} - \Delta{\left(r \right)}\right) \sin^{2}{\left(\theta \right)}}{M \Sigma{\left(r,\theta \right)}}\\0 & \frac{\Sigma{\left(r,\theta \right)}}{\Delta{\left(r \right)}} & 0 & 0\\0 & 0 & \Sigma{\left(r,\theta \right)} & 0\\- \frac{J \left(\frac{J^{2}}{M^{2}} + r^{2} - \Delta{\left(r \right)}\right) \sin^{2}{\left(\theta \right)}}{M \Sigma{\left(r,\theta \right)}} & 0 & 0 & \frac{\left(- \frac{J^{2} \Delta{\left(r \right)} \sin^{2}{\left(\theta \right)}}{M^{2}} + \left(\frac{J^{2}}{M^{2}} + r^{2}\right)^{2}\right) \sin^{2}{\left(\theta \right)}}{\Sigma{\left(r,\theta \right)}}\end{matrix}\right] \\
A^{\text{KN}}_\mu = \left[\begin{matrix}\frac{Q r}{\Sigma{\left(r,\theta \right)}}\\0\\0\\- \frac{J Q r \sin^{2}{\left(\theta \right)}}{M \Sigma{\left(r,\theta \right)}}\end{matrix}\right]\\
\end{gathered}
\label{KN sol}
\end{equation}
with
\begin{equation}
\begin{split}
    \Sigma(r,\theta) &= \frac{J^{2} \cos^{2}{\left(\theta \right)}}{M^{2}} + r^{2} \\
    \Delta(r,\theta) &= \frac{J^{2}}{M^{2}} - 2 M r + Q^{2} + r^{2}\;,
\end{split}
\end{equation}
\twocolumngrid
\noindent into the NC Einstein and NC Maxwell equations we find the $a^1$ source terms for the $h_{\mu\nu}$ and $B_\mu$ as the evaluation of the NC equations \eqref{EFE} and \eqref{Maxwell} on the commutative Kerr-Newmann solution, i.e., as the part of the equations which the KN electrovacuum \eqref{KN sol} does not solve. In the NC Maxwell equation \eqref{Maxwell}, $\lambda = r,\varphi,\theta$ equations are not solved by the KN commutative ansatz and there are around 50 terms, given as rational functions, in each nonzero component. For the NC Einstein equation we find that there are only 4 simple distinct source term components arising from plugging in the Kerr-Newmann solution into \eqref{EFE}:
\widetext
\begin{equation}
    G_{\mu\nu}(g^{\text{KN}}) - 2T_{\mu\nu}(g^{\text{KN}}) = aS_{\mu\nu} = \left[\begin{matrix}0 & aS_{tr} & aS_{t\theta} & 0\\aS_{rt} & 0 & 0 & aS_{r\varphi}\\aS_{\theta t} & 0 & 0 & aS_{\theta\varphi}\\0 & aS_{\varphi r} & aS_{\varphi\theta} & 0\end{matrix}\right]
\label{Smunu}
\end{equation}
with
\begin{equation}
\begin{gathered}
    S_{tr} = S_{rt} = - \frac{4 J Q^{3} r^{2} \left(r \frac{\partial}{\partial r} \Sigma{\left(r,\theta \right)} - \Sigma{\left(r,\theta \right)}\right) \sin{\left(2 \theta \right)} \frac{\partial}{\partial \theta} \Sigma{\left(r,\theta \right)}}{M \Sigma^{6}{\left(r,\theta \right)}}\;, \\
    S_{r\varphi} = S_{\varphi r} = \frac{8 J^{2} Q^{3} r^{2} \left(r \frac{\partial}{\partial r} \Sigma{\left(r,\theta \right)} - \Sigma{\left(r,\theta \right)}\right) \left(- 2 \Sigma{\left(r,\theta \right)} \cos{\left(\theta \right)} + \sin{\left(\theta \right)} \frac{\partial}{\partial \theta} \Sigma{\left(r,\theta \right)}\right) \sin^{2}{\left(\theta \right)} \cos{\left(\theta \right)}}{M^{2} \Sigma^{6}{\left(r,\theta \right)}}\;,\\
    S_{t\theta} = S_{\theta t} = \frac{4 J Q^{3}r \left(r \frac{\partial}{\partial r} \Sigma{\left(r,\theta \right)} - \Sigma{\left(r,\theta \right)}\right)^{2} \Delta{\left(r \right)} \sin{\left(2 \theta \right)}}{M \Sigma^{6}{\left(r,\theta \right)}}\;,\\
    S_{\varphi\theta} = S_{\theta\varphi} = - \frac{8 J^{2} Q^{3}r \left(r \frac{\partial}{\partial r} \Sigma{\left(r,\theta \right)} - \Sigma{\left(r,\theta \right)}\right)^{2} \Delta{\left(r \right)} \sin^{3}{\left(\theta \right)} \cos{\left(\theta \right)}}{M^{2} \Sigma^{6}{\left(r,\theta \right)}}\;.
\end{gathered}
\end{equation}
\twocolumngrid
\section{Solving the NC Einstein-Maxwell equation}
By inspecting the form of \eqref{Smunu} one is inclined to try the ansatz for the noncommutative part of the metric with 4 nonzero components
\begin{equation}
h_{\mu\nu} = \left[\begin{matrix}0 & h_{tr}(r,\theta) & h_{t\theta}(r,\theta) & 0\\h_{tr}(r,\theta) & 0 & 0 &h_{r\varphi}(r,\theta)\\h_{t\theta}(r,\theta) & 0 & 0 & h_{\theta\varphi}(r,\theta)\\0 & h_{r\varphi}(r,\theta) & h_{\theta\varphi}(r,\theta) & 0\end{matrix}\right]\;,
\label{general hmunu ansatz}
\end{equation}
but simply using \eqref{general hmunu ansatz} as the $a^1$ part of the ansatz \eqref{ansatz} and plugging it in \eqref{EFE} will give a system of 4 partial differential equations with each equation being hundreds of terms long (without even yet considering the Maxwell part). So solving the system of equations in general is computationally intractable \footnote{Calculating the system of equations for the ansatz \eqref{general hmunu ansatz} took two hours on a modern PC using the open source Sympy library.} unless one specifies the ansatz in much more detail. The way the author has found the solution was by using physical arguments, not by mechanical computation.\\

Firstly, the electromagnetic part of the solution should be the Seiberg-Witten map of the standard Kerr-Newman potential
\begin{equation}
    A_\mu = \left[\begin{matrix}\frac{Q r}{\Sigma{\left(r,\theta \right)}}\\0\\0\\- \frac{J Q r \sin^{2}{\left(\theta \right)}}{M \Sigma{\left(r,\theta \right)}}\end{matrix}\right] +a\left[\begin{matrix}0\\0\\- \frac{qJ Q^{2}r^{2} \sin{\left(2 \theta \right)}}{2 M \Sigma^{2}{\left(r,\theta \right)}}\\0\end{matrix}\right]\;,
\label{swamu}
\end{equation}
which is not surprising since the theory that we are solving is a $U(1)_\star$ noncommutative gauge theory. The metric part of the solution can be obtained by considering the scalar field action \eqref{Andelo action} in the (commutative) KN background, and in the manner of \cite{DimitrijevicCiric:2024ibc}, the author searched for an effective metric $\hat{g}$ for which the Klein-Gordon operator equals the NC scalar field's equation of motion in the fixed KN background
\begin{equation}
    \frac{\delta S}{\delta \phi^\dagger} = \square_{\hat{g}}\phi\;.
\label{effective}
\end{equation}
To find the solution $\hat{g}$ of the equation \eqref{effective}, the author used the metric part of the ansatz \eqref{ansatz}
\begin{equation}
    \hat{g}_{\mu\nu} = g^{\text{KN}}_{\mu\nu} + a\hat{h}_{\mu\nu}\;,
\end{equation}
and by comparing second-derivative terms of $\phi$ in the right-hand and the left-hand side of \eqref{effective}, the unknown metric components were found to be the variables of an algebraic system of equation whose solutions are
\begin{equation}
\begin{split}
\hat{h}_{tr} &= - \frac{qJ M Q \left(J^{2} \cos^{2}{\left(\theta \right)} - M^{2} r^{2}\right) \sin^{2}{\left(\theta \right)}}{2 \left(J^{2} \cos^{2}{\left(\theta \right)} + M^{2} r^{2}\right)^{2}}\;, \\
\hat{h}_{t\theta} &= - \frac{qJ M^{3} Q r \Delta{\left(r \right)} \sin{\left(2 \theta \right)}}{2 \left(J^{2} \cos^{2}{\left(\theta \right)} + M^{2} r^{2}\right)^{2}}\;,\\
\hat{h}_{r\varphi} &= \frac{qQ \left(J^{2} + M^{2} r^{2}\right) \left(J^{2} \cos^{2}{\left(\theta \right)} - M^{2} r^{2}\right) \sin^{2}{\left(\theta \right)}}{2 \left(J^{2} \cos^{2}{\left(\theta \right)} + M^{2} r^{2}\right)^{2}}\;,\\
\hat{h}_{\theta\varphi} &= \frac{qJ^{2} M^{2} Q r \Delta{\left(r \right)} \sin^{2}{\left(\theta \right)} \sin{\left(2 \theta \right)}}{2 \left(J^{2} \cos^{2}{\left(\theta \right)} + M^{2} r^{2}\right)^{2}}\;.\\
\end{split}
\label{hhat components}
\end{equation}
 The effective metric \eqref{effective} for the scalar field reorganizes the NC gauge symmetry of the scalar field and electromagnetic background into the metric effects, but without considering the back-reaction of the gravitational field. In the problem we are solving, the SW expanded $U(1)_\star$ potential again drives the reaction and back-reaction. With that said, one can try to see if this back-reaction is of the same form as the effective metric. Following that logic, let us use the effective metric \eqref{hhat components} as an ansatz for the Einstein-Maxwell equations \eqref{EFE}, \eqref{Maxwell} by allowing the numerical coefficients of $\hat{h}_{\mu\nu}$ to vary. This amounts to setting the $h_{\mu\nu}$ field in the ansatz \eqref{ansatz} to
\widetext
\begin{equation}
    h_{\mu\nu} = \left[\begin{matrix}0 & C_1 \hat{h}_{tr}(r,\theta) & C_2 \hat{h}_{t\theta}(r,\theta) & 0\\ C_1 \hat{h}_{tr}(r,\theta) & 0 & 0 &C_3 \hat{h}_{r\varphi}(r,\theta)\\ C_2 \hat{h}_{t\theta}(r,\theta) & 0 & 0 & C_4 \hat{h}_{\theta\varphi}(r,\theta)\\0 & C_3 \hat{h}_{r\varphi}(r,\theta) & C_4 \hat{h}_{\theta\varphi}(r,\theta) & 0\end{matrix}\right]\;.
\label{hmunu ansatz}
\end{equation}
Incredibly, the ansatz turns out to be functionally correct. This can be seen by plugging it into the Einstein-Maxwell problem \eqref{EFE}, \eqref{Maxwell} which in turn uniquely imposes setting all of the unknown coefficients to
\begin{equation}
    C_1=C_2=C_3=C_4 = -2\;.
\label{coeffs}
\end{equation}
The ansatz \eqref{hmunu ansatz} (with the restriction \eqref{coeffs}) together with the Seiberg-Witten expanded potential \eqref{swamu} solves the NC equations of motion. In conclusion, a charged rotating solution (up to $\mathcal{O}(a^2)$) of the NC Einstein-Hilbert-Maxwell action \eqref{NC EHM} is given as
\begin{equation}
\begin{gathered}
    g_{\mu\nu} = \left[\begin{matrix}\frac{\frac{J^{2} \sin^{2}{\left(\theta \right)}}{M^{2}} - \Delta{\left(r \right)}}{\Sigma{\left(r,\theta \right)}} & -2a \hat{h}_{tr}{\left(r,\theta \right)} & -2a\hat{h}_{t\theta}{\left(r,\theta \right)} & - \frac{J \left(\frac{J^{2}}{M^{2}} + r^{2} - \Delta{\left(r \right)}\right) \sin^{2}{\left(\theta \right)}}{M \Sigma{\left(r,\theta \right)}}\\-2a\hat{h}_{tr}{\left(r,\theta \right)} & \frac{\Sigma{\left(r,\theta \right)}}{\Delta{\left(r \right)}} & 0 & -2a\hat{h}_{r\varphi}{\left(r,\theta \right)}\\-2a\hat{h}_{t\theta}{\left(r,\theta \right)} & 0 & \Sigma{\left(r,\theta \right)} & -2a\hat{h}_{\theta\varphi}{\left(r,\theta \right)}\\- \frac{J \left(\frac{J^{2}}{M^{2}} + r^{2} - \Delta{\left(r \right)}\right) \sin^{2}{\left(\theta \right)}}{M \Sigma{\left(r,\theta \right)}} & -2\hat{h}_{r\varphi}{\left(r,\theta \right)} & -2a\hat{h}_{\theta\varphi}{\left(r,\theta \right)} & \frac{\left(- \frac{J^{2} \Delta{\left(r \right)} \sin^{2}{\left(\theta \right)}}{M^{2}} + \left(\frac{J^{2}}{M^{2}} + r^{2}\right)^{2}\right) \sin^{2}{\left(\theta \right)}}{\Sigma{\left(r,\theta \right)}}\end{matrix}\right]\;,\\
    A_\mu = \left[\begin{matrix}\frac{Q r}{\Sigma{\left(r,\theta \right)}}\\0\\- a\frac{qJ Q^{2} r^{2} \sin{\left(2 \theta \right)}}{2 M \Sigma^{2}{\left(r,\theta \right)}}\\- \frac{J Q r \sin^{2}{\left(\theta \right)}}{M \Sigma{\left(r,\theta \right)}}\end{matrix}\right]\;,
\end{gathered}
\label{solution}
\end{equation}
\twocolumngrid
\noindent and it satisfies the condition that $\partial_t$ and $\partial_\varphi$ are Killing vectors for $g_{\mu\nu}$ and $A_\mu$ so our solution \eqref{solution} is valid for any action of the form \eqref{NC action} which reproduces the usual Einstein-Hilbert-Maxwell action in the limit \footnote{In other words, the solution \eqref{solution} is valid for any action which is obtained from the Einstein-Hilbert Maxwell action by introducing the NC geometry and NC gauge theory as outlined in earlies Sections. As discussed in \eqref{deformations}, a very big space of NC Lagrangians is compatible with this framework.} $a \rightarrow 0$.\\

It is worth spending a few words on some properties of the noncommutative solution \eqref{solution}. Firstly, its Komar mass and angular momentum are $M$ and $J$ respectively, so they remain unchanged in the noncommutative spacetime. Secondly, we can see that the dimensional parameter $q$ is present in both the metric and potential part of the solution. This dimensional parameter equals the charge of the scalar field in the theory \eqref{Andelo action}, but since our theory \eqref{NC EHM} does not have a scalar field, we can not decide on a specific value of the parameter $q$. Whether the parameter $q$ is something fundamental like a coupling constant, or if it is a new type of black hole hair, from the presented analyses it is not clear \footnote{It is not unreasonable to imagine that two NC Kerr-Newman solutions could model physical reality with the same commutative hairs $M,J,Q$ but with a different $q$. On the other hand, it also seems plausible that the same $q$ applies to all modifications of commutative solutions, which is already the case for the parameter $a$.}. Notice that in the Seiberg-Witten map \eqref{SW map} the presence of the parameter $q$ is necessary because otherwise the map would not be dimensionally consistent. One thing is certain, for any value of $q$ we obtain a solution to the NC Einstein Maxwell problem. \\

Finally, in the limit $J\rightarrow 0$ we obtain the same metric (up to the already mentioned $-2$ factor) as the effective NC Reissner–Nordström metric in \cite{DimitrijevicCiric:2024ibc} where it was also noticed that the effective metric and commutative electromagnetic potential solve the NC Einstein-Maxwell equation \footnote{In the limit $J\rightarrow 0$ the electromagnetic potential is simple enough that it does not introduce any $a^1$ terms to the Einstein equation or the Maxwell equation. The NC Einstein-Maxwell problem in that case is the same as the commutative one.} up to $\mathcal{O}(a^2)$.
\section{Final remarks}
In this letter we have explored the inclusion of noncommutative geometry via the $\partial_t\wedge\partial_\varphi$ coordinate Drinfeld twist into the standard Einstein-Hilbert-Maxwell action. In the discussion around \eqref{deformations}, we have concluded that, for every $\theta^{\mu\nu},$ there are infinitely many equally motivated models of minimally NC deformed Einstein-Maxwell theory. This means that, unless one finds a solution which they all share (for a given $\theta^{\mu\nu}$), exploring physical consequences of such deformations would be troublesome. Afterwards, we have seen that for the $\partial_t\wedge\partial_\varphi$ twist, all of the minimal NC deformations \eqref{NC action} of the Einstein-Maxwell theory share a solution in case there exists a $\partial_t,\partial_\varphi$ Killing symmetric solution to the equations of motion of \eqref{NC EHM}. In case such a solution to the EOM of \eqref{NC EHM} exists, it will also satisfy the EOM for all actions \eqref{NC action}. The author would like to emphasize how big this space of minimally NC deformed Einstein-Maxwell actions is. For example, in the sense of \eqref{deformations}, the reader is welcome to expand the commutative Ricci scalar and to count how many inequivalent orderings of primitive objects there can be upon promoting the commutative $\cdot$ product to the $\star$ product. Additionally, a weighted average of such deformations is also a valid NC deformation. Never the less, all such NC deformations (in the $\partial_t\wedge\partial_\varphi$ twist) of the Einstein-Hilbert action along with any NC deformation of the Maxwell action will share the solution \eqref{solution} to the equations of motion.\\

The solution \eqref{solution} is actually a family of solutions, parametrized by $q$, which correspond to perturbative corrections to the commutative Kerr-Newman solution, with 4 new metric components, $g_{tr}, g_{t\theta}, g_{r\varphi}, g_{\varphi\theta}$ as well as one new electromagnetic potential component, $A_\theta$, appearing in the solutions. Whether the parameter $q$ is more realistically described as some noncommutative hair of the NC Kerr-Newman black hole, or rather as a constant in the model, is currently unresolved. This letter is the first time, to author's awareness, that the $\partial_t\wedge\partial_\varphi$ Drinfeld twist deformation formalism was applied to rotating black hole physics. It is also the first time that in this formalism a correction to the black hole physics is obtained directly as a solution to the equation of motion of the NC action and not by some indirect methods. This letter serves as the proof of concept that the quantum nature of spacetime can indeed affect the black hole solutions. Finally, \cite{Aschieri:2009qh} (and further applied in \cite{Ohl:2009pv}) established a result that commutative gravitational solutions with (at least) two Killing vectors are also solutions to the noncommutative Einstein equation constructed from a Killing Drinfeld twist. Here we give more depth to this result by proving that even if the vectors that define the twist are Killing for all of the commutative fields, they can still fail to solve the noncommutative Einstein-matter equation in case the matter content contains gauge fields complicated enough that the Seiberg-Witten map can introduce new nonzero source terms. Never the less, the claim from \cite{Aschieri:2009qh} can be slightly weakened to say that in such cases, the left-hand side of the Einstein equation remains undeformed, it is just the sources which can be deformed from the deformation of gauge symmetry.
\section{Acknowledgment}
This research was supported by the Croatian Science Foundation Project IP-2020-02-9614 ”Search for Quantum spacetime in Black Hole QNM spectrum and Gamma Ray Bursts”. The author would like to thank dr. Naveena Kumara Athithamoole for useful discussion.

\bibliography{main}

\end{document}